\documentclass[aps,preprint,twoside,nobalancelastpage,12pt,a4paper,onecolumn,secnumarabic]{revtex4}
\usepackage{amsmath}
\usepackage{amssymb}
\usepackage{amsfonts}
\usepackage{graphicx}
\begin{document}
\title{Self-organized multiscale structures in thermally relativistic
electron-positron-ion plasmas}
\author{Usman Shazad}
\email{usmangondle@gmail.com}
\author{M. Iqbal}
\author{Shafa Ullah}
\affiliation{Department of Physics, University of Engineering and Technology, Lahore
54890, Pakistan}

\begin{abstract}
The self-organization of a thermally relativistic magnetized plasma
comprising of electrons, positrons and static ions is investigated. The
self-organized state is found to be the superposition of three distinct
Beltrami fields known as triple Beltrami (TB) state. In general, the
eigenvalues associated with the multiscale self-organized vortices may be a
pair of complex conjugate and real one. It is shown that all the eigenvalues
become real when thermal energy increases or~the positron density decreases.
The impact of relativistic temperature and positron density on the formation
of self-organized structures is investigated. The self-organized field and
flow vortices may vary simultaneously on vastly different length scales. The
disparate variation of self-organized vortices is important in the context
of dynamo theory. The present work is useful to study the formation of
multiscale vortices and dynamo mechanisms in multi-species thermally
relativistic plasmas.
\end{abstract}

\maketitle

\section{Introduction}

The self-organization occurs throughout the universe. The magnetized plasmas
also self-organize themselves. The self-organization in plasmas is also
called relaxation \cite{Hasegawa1986}. In magnetized plasmas, the
self-organization minimizes the magnetofluid energy under certain
constraints and lead the turbulent (disordered) state towards an equilibrium
(ordered) state. The magnetohydrodynamic (MHD) plasmas self-organize to
force-free state called as Beltrami state. Mathematically the Beltrami state
is the Euler-Lagrange equation and expressed by an eigenvalue equation of
the curl operator. The magnetic field acts as an eigenfunction and satisfies
the relation $\mathbf{\nabla }\times \mathbf{B}=\mu \mathbf{B}$, where the
eigenvalue $\mu $ is a constant and represents the ratio of current to the
magnetic field \cite{Chandrasekhar1958}. The eigenvalue equation was derived
by Woltjer and Taylor using variational principle and called as
Woltjer-Taylor state \cite{Woltjer1958,Taylor1974,Taylor1986}.

The MHD model of self-organization was extended to Hall MHD (HMHD) plasma to
incorporate the missing features in MHD like pressure gradients and flows 
\cite{Steinhauer1997,Mahajan1998}. The self-organized state of HMHD plasma
is a non force-free state and can be expressed as a combination of\ two
different Beltrami states called double Beltrami state (DB). The salient
features of DB states are the strong coupling of the magnetic field and
flow, high beta, diamagnetism and self-confinement of plasmas \cite%
{Mahajan1998,Steinhauer1998,Steinhauer2002}. The DB states have been
extensively used to model fusion \cite%
{Yoshida2001,Lorenzini2009,Steinhauer2011,Jardin2015} and astrophysical
plasmas such as flow generation in solar atmosphere and compact
astrophysical objects \cite{Mahajan2002,Barnaveli2017}, dynamo and reverse
dynamo mechanisms \cite{Mahajan2005}, multi-scale structure formation in
space and astrophysical plasmas \cite{Dasgupta2009}, formation of solar
arcades and coronal mass ejection \cite{Ohsaki2002,Kagan2010}, and
diamagnetic states in cosmological plasmas \cite{Asenjo2019}. The inertia of
the plasma species also plays a very important role in the process of
self-organiztion and introduces coupling of multiple Beltrami states. For
instance, when the of inertial effects of both the plasma componets are
taking into account, the relaxed state comes out to be a Triple Beltrami
(TB) state - a superposition of three Beltrami states \cite{Taveira2003}.

The study of self-organization in thermally relativistic plasmas has
attracted the attention of many researchers. The thermally relativistic
plasmas exhibit temperature which is higher than the rest mass energy of
electrons. The study of relativistic thermal plasmas is of potential
importance in the fields of laser plasma interaction and high energy
astrophysics. Several theoretical and simulational studies suggest that
thermally relativistic plasmas of multi-MeV temperatures can quickly be
generated by intense laser pulses \cite{Shukla2005,Li2008,Li2011}The data
obtained by high energy X-rays and $\gamma $-rays provide a strong evidence
that there exist places in the universe with temperatures higher than 1 MeV 
\cite{Lightman1982}. The time up to one second after the big bang, the
temperature of the early universe was in the MeV range. The core
constituents of the universe during this time were electrons and positrons 
\cite{Rees1984}. In present day epoch, the thermally relativistic electron
and positron (EP) plasmas are believed to occur in pulsar magnetosphere \cite%
{Goldreich1969}, active galactic nuclei (AGN) \cite{Lightman1987}, hot
accretion disks of black holes \cite{Takahara1985}, M87 jet \cite%
{Reynolds1996} and galactic center of our galaxy \cite{Cordier2004}. The EP
plasmas can naturally coexist with the ion species which are ubiquitous in
astrophysical environments \cite%
{Begelman1984,Rizzato1988,Holcomb1989,Berezhiani1994,Berezhiani1995,Shatashvili2016}%
.

It was shown by Iqbal et al that a thermally relativistic EP plasma can be
self-organized to a TB state. Furthermore, it has been shown that the
relativistic temperature controls the size of self-organized structures \cite%
{Iqbal2008}. The relaxed state is likewise a TB state for a two-temperature
thermally relativistic EPI. In this plasma model, the positron density is
considered to be negligible. It has been discovered that when the
relativistic temperature increases, the eigenvalues become complex \cite%
{Iqbal2012,Iqbal2013}. In another study based on the minimum fluid coupling
model, the relaxation of a relativistically hot plasma is studied and the
applicability of the results to astrphysical phenomena (the striped wind\ of
a pulsar nebula) have been discussed \cite{Pino2010}.

Recently, it has been studied that electron degeneracy pressure may enable a
new kind of Beltrami-Bernoulli (BB) equilibrium for a dense degenerate
electron-ion plasma. These states are theoretically investigated for new
energy transformations, such as degeneracy energy into fluid kinetic energy
and are very important for understanding of white dwarfs and neutron stars\ 
\cite{Berezhiani2015}. The relaxed state of relativistic degenerate EPI
plasma, which is composed of degenerate electrons and positrons with a small
fraction of mobile classical ions is found to be a Quadruple Beltrami (QB)
state. It is demonstrated that increased effective inertia of bulk EP
components as a result of temperature and degeneracy increases effective
skin depths, and that ion contamination contributes to the development of
intermediate and macro scales by enriching structure formation and expanding
energy transformation pathways\ \cite{Shatashvili2016}. In a more recent
study, Shatashvili et al., investigated the quasi equilibrium
Beltrami-Bernoulli states of a three-component plasma consisting of two
electron species immersed in a neutralising ion background. Furthermore, it
has been shown that the QB state is the relaxed state for this plasma system 
\cite{Shatashvili2019}.

The present work is devoted to explore the possibility of self-organized
state of thermally relativistic EPI plasma to TB state. We assume an
incompressible and quasi-neutral thermally relativistic
electron-positron-ion plasma. The positive ions are taken to be stationary.
The positive ions break the symmetry and play the role to keep the plasma as
quasi-neutral. The electrons and positrons are supposed to be thermally
relativistic so that their thermal energy is greater than or equal to their
rest mass energies. However, the directed velocity of the plasma is
considered to be non-relativistic.\ It is shown that the system
self-organizes to TB state. The analysis shows that for lower relativistic
temperature and higher positron density, the scale parameters are complex
but with an increase in thermal energy and lower positron density, the scale
parameters become real. It is shown that paramagnetic structures can be
transformed to diamagnetic ones or vice-versa on varying the temperatures
and densities of species. Such a transformation of magnetic field is
important in the understanding of magnetic reconnection (which contributes
in heating and cooling of plasma) and generation of fast outflows. It is
also shown that for appropriate Beltrami parameters, it is possible to
create self-organized field and flow vortices varying on different length
scales that can manifest the dynamo mechanisms in TB state.

The manuscript is arranged as follows. Essential equations for the plasma
system are delineated and a TB state is obtained in Sec. 2. In Sec. 3, the
variational principle approach is used to derive TB state. Sec. 4 is devoted
to describe the characteristics of scale parameters and the impact of
thermal energy and positron density on them. The analytical solution of TB
equation is presented and how the positron density and thermal energy effect
the self-organized process is discussed in Sec. 5. The field and flow
profiles showing the dynamo mechanisms are described in Sec. 6. The summary
of the work is presented in Sec. 7.

\section{Model Equations and TB state}

We consider a three component incompressible and collisionless plasma. The
components are electrons, positrons and ions. The ions are static while the
electrons and positrons are thermally relativistic.\ When the directed fluid
velocity approaches the speed of light, the plasma is called relativistic.
It is also referred to be relativistic when the thermal energy of the
components is equal to or greater than their rest mass energy. Both types of
relativistic plasmas are encountered in astrophysical and laboratory
settings. Laboratory relativistic plasmas may be generated and accelerated
using intense laser pulses.\ In the present work, the word relativistic is
used for electrons and positrons whose thermal energy is greater than or
equal to their rest mass energy.\ For the velocity distribution of the
particles to be a local relativistic Maxwellian, the factor $G(z_{\alpha
})=K_{3}(1/z_{\alpha })/K_{2}(1/z_{\alpha })$ shows the effect of
relativistic temperature or thermal energies of plasma species. In the
factor $G(z_{\alpha }),$ $K_{2}$ and $K_{3}$ are modified Bessel functions
of order 2 and 3\ respectively and $z_{\alpha }=T_{\alpha }/m_{0\alpha
}c^{2} $ where $m_{0\alpha }$ and $T_{\alpha }$ are the invariant rest
masses and temperatures of the particles respectively. The factor $%
G(z_{\alpha })$ has the following asypmtotic approximations: when the
thermal energy is less than rest mass energy of plasma species $z_{\alpha
}<<1,$ the plasma is in non-relativistic regime and $G(z_{\alpha })\approx
1+5z_{\alpha }/2$ but for highly relativistic plasma, $z_{\alpha }>>1$ and $%
G(z_{\alpha })\approx 4z_{\alpha }$ \cite{Berezhiani1995}. The
quasi-neutrality condition reads as%
\begin{equation}
N_{p}+N_{i}=1,  \label{qn}
\end{equation}%
where $N_{p}=n_{p}/n_{e}$ and $N_{i}=n_{i}/n_{e}$ in which $n_{e}$, $n_{p}$
and $n_{i}\ $are number densities of electrons, positrons and ions,
respectively. By following the Ref. \cite{Berezhiani2002}, the equations of
motion for thermally relativistic electrons and positrons can be expressed as%
\begin{equation}
\frac{\partial }{\partial t}\left( G_{\alpha }m_{0\alpha }\gamma _{\alpha }%
\mathbf{V}_{\alpha }\right) +m_{o\alpha }c^{2}\mathbf{\nabla }\left(
G_{\alpha }\gamma _{\alpha }\right) =q_{\alpha }\mathbf{E}+\mathbf{V}%
_{\alpha }\times \mathbf{\Omega }_{\alpha },  \label{me}
\end{equation}%
where the index $\alpha $ equals `$p$' for positrons and `$e$' for electrons
and $\mathbf{\Omega }_{\alpha }=\mathbf{\nabla }\times \left( G_{\alpha
}m_{0\alpha }\gamma _{\alpha }\mathbf{V}_{\alpha }\right) +q_{\alpha }%
\mathbf{B/}c$. $G_{\alpha }$, $m_{0\alpha }$, $\gamma _{\alpha }$, $\mathbf{V%
}_{\alpha }$, $q_{\alpha }$, $c$, $\mathbf{E}$, and $\mathbf{C}$ represent
relativistic temperature, rest mass, relativistic Lorentz factor, velocity,
charge, speed of light, electric field, and magnetic field respectively. The
magnetic and electric fields are related to vector potential $\left( \mathbf{%
A}\right) $ and scalar potential $\left( \phi \right) $ by the relations $%
\mathbf{B}$ $=\mathbf{\nabla }\times \mathbf{A}$ and $\mathbf{E=-\nabla }%
\phi -c^{-1}\partial \mathbf{A}/\partial t$, respectively. The electrons and
positrons are antiparticles, so their masses are equal $\left(
m_{0e}=m_{0p}\right) $ and oppositely charged $\left( q_{e}=-e\text{ and }%
q_{p}=e\right) $. Only the electrons and positrons are taken to be thermally
relativistic while the directed fluid velocity $\mathbf{V}_{\alpha }<<c,$ so
the relativistic Lorentz factor becomes $\gamma _{\alpha }=\left( 1-\mathbf{V%
}_{\alpha }^{2}/c^{2}\right) ^{-1/2}\approx 1$. The term $m_{o\alpha }c^{2}%
\mathbf{\nabla }\left( G_{\alpha }\gamma _{\alpha }\right) $ in Eq. (\ref{me}%
) accounts for pressure gradient. The relation between thermal pressure $%
p_{\alpha }$ and relativistic temperature $G_{\alpha }$ is $\gamma _{\alpha
}\nabla p_{\alpha }=m_{0\alpha }c^{2}n_{\alpha }\mathbf{\nabla }\left(
G_{\alpha }\right) $, where $p_{\alpha }=n_{\alpha }T_{\alpha }/\gamma
_{\alpha }$. The Eq. (\ref{me}) is augmented by the following equation of
state%
\begin{equation}
\frac{n_{\alpha }}{\gamma _{\alpha }}\frac{z_{\alpha }}{K_{2}\left(
z_{\alpha }\right) }\exp \left( -z_{\alpha }G_{\alpha }\right) =\text{%
constant}.  \label{State}
\end{equation}%
The plasma pressure is considered isotropic and for simplicity we assume
that the relativistic temperatures of electrons and positrons are equal, $%
G_{e}=G_{p}=G$. To express Eq. (\ref{me}) in dimensionless form, all the
lengths are normalized by electron skin depth $\lambda _{e}$ and time with
inverse of electron plasma frequency $\omega _{pe}$, where $\lambda _{e}=%
\sqrt{m_{0e}c^{2}\left( 4\pi n_{e}e^{2}\right) ^{-1}}$ and $\omega _{pe}=%
\sqrt{4\pi n_{e}e^{2}m_{0e}^{-1}}.$ The magnetic field $B$, flows $V_{\alpha
}$ and pressure term $\left( m_{0\alpha }c^{2}\nabla G_{\alpha }\right) $
are normalized with\ some arbitrary value of magnetic field $B_{0}$, Alfv%
\'{e}n velocity $V_{A}=B_{0}/\sqrt{4\pi m_{0e}n_{e}}$ and $B_{0}^{2}/\left(
4\pi n_{e}m_{0e}c^{2}\right) ^{-1}$ respectively. To obtain the vortex
dynamic equations, we take the curl of Eq. (\ref{me}). The vortex dynamic
equations are%
\begin{equation}
\frac{\partial \mathbf{\Omega }_{\alpha }}{\partial t}=\mathbf{\nabla }%
\times \lbrack \mathbf{V}_{\alpha }\times \mathbf{\Omega }_{\alpha }],
\label{ve}
\end{equation}%
where $\mathbf{\Omega }_{\alpha }=\mathbf{\nabla }\times G\mathbf{V}_{\alpha
}+q_{\alpha }\mathbf{B}$\textbf{\ }is the generalized or canonical
vorticity. It is easy to show that when the gradient forces ($\mathbf{\nabla 
}\psi _{j}=\mathbf{\nabla }G_{j}+q_{j}\mathbf{\nabla }\phi _{j}$) are
considered to be zero individually, the relaxed state with the constraint $%
\mathbf{V}_{j}\times \mathbf{\Omega }_{j}=0$ defines an equilibrium state.
Although the latter has generalized Bernoulli conditions ($\psi _{j}$%
=constant), however these are irrelevant to the analysis presented in this
article. The steady-state solution of Eq. (\ref{ve}) yields the two Beltrami
conditions for electrons and positrons as follows%
\begin{equation}
\mathbf{\nabla }\times G\mathbf{V}_{e}-\mathbf{B}=aG\mathbf{V}_{e},
\label{be}
\end{equation}%
\begin{equation}
\mathbf{\nabla }\times G\mathbf{V}_{p}+\mathbf{B}=bG\mathbf{V}_{p},
\label{bp}
\end{equation}%
where $a$ and $b$ are the Beltrami parameters for electrons and positrons
respectively.\textbf{\ }The Beltrami parameters $a$ and $b$ are ratios of
generalized vorticities to their respective flows.\textbf{\ }The Beltrami
conditions for plasma species describe their independent dynamics. To couple
the dynamics of plasma species, Ampere's law is adopted. For this plasma
system, Ampere's law in dimensionless form is%
\begin{equation}
\mathbf{\nabla }\times \mathbf{B}=N_{p}\mathbf{V}_{p}-\mathbf{V}_{e}.
\label{al}
\end{equation}%
The Eqs. (\ref{be}-\ref{al}) will be employed to derive a relaxed state for
the plasma system. Eliminating $\mathbf{V}_{e}$ from Eqs. (\ref{be}) and (%
\ref{al}), $\mathbf{\mathbf{V}_{p}}$ is obtained%
\begin{equation}
\mathbf{\mathbf{V}_{p}}=\frac{1}{N_{p}(b-a\mathbf{)}}[\mathbf{\mathbf{\nabla 
}\times \mathbf{\nabla }\times \mathbf{B}-}a\mathbf{\mathbf{\nabla }\times 
\mathbf{B+(}}\frac{1+N_{p}}{G}\mathbf{\mathbf{)B]}},  \label{vp}
\end{equation}%
Using Eqs. (\ref{bp}) and (\ref{vp}), we obtain%
\begin{equation}
\mathbf{\mathbf{\nabla }\times \mathbf{\nabla }\times \mathbf{\nabla }\times 
\mathbf{\mathbf{B}-}}k_{3}\mathbf{\mathbf{\nabla }\times \mathbf{\nabla
\times \mathbf{B+}}}k_{2}\mathbf{\mathbf{\mathbf{\mathbf{\nabla }\times B}}}%
-k_{1}\mathbf{B}=0.  \label{tcb}
\end{equation}%
where $k_{1}=\left( b+aN_{p}\right) /G$, $k_{2}=\left( 1+N_{p}\right) /G+ab$
and $k_{3}=a+b$. Eq. (\ref{tcb}) is the steady state equilibrium (relaxed)
state and known as TB equation. In evaluating eq. (\ref{tcb}) all the linear
and non-linear effects are taken into the account.

To find the composite flow $\mathbf{V}$, we first find expressions of
electrons and positrons flows. The positron velocity using Eq. (\ref{vp})
can be written as 
\begin{equation}
\mathbf{\mathbf{V}_{p}}=p_{3}\mathbf{\mathbf{\nabla }\times \mathbf{\nabla }%
\times \mathbf{B}-}p_{2}\mathbf{\mathbf{\nabla }\times \mathbf{B+}}p_{1}%
\mathbf{\mathbf{B,}}  \label{vp1}
\end{equation}%
where $p_{3}=\left[ N_{p}(b-a\mathbf{)}\right] ^{-1}$, $p_{2}=a\left[
N_{p}(b-a\mathbf{)}\right] ^{-1}$ and $p_{1}=\mathbf{\mathbf{(}}1+N_{p}%
\mathbf{\mathbf{)}}\left[ GN_{p}(b-a\mathbf{)}\right] ^{-1}$\textbf{.} Using
Eq. (\ref{vp1}) in Eq. (\ref{al}), we get the electron velocity $\mathbf{V}%
_{e}\mathbf{\ }$as given by%
\begin{equation}
\mathbf{V}_{e}=e_{3}\mathbf{\mathbf{\nabla }\times \mathbf{\nabla }\times 
\mathbf{\mathbf{B}-}}e_{2}\mathbf{\mathbf{\mathbf{\nabla }\times \mathbf{B+}}%
}e_{1}\mathbf{\mathbf{\mathbf{B}},}  \label{ve1}
\end{equation}%
where $e_{3}=p_{3}N_{p}$, $e_{2}=p_{2}N_{p}+1$ and $e_{1}=p_{1}N_{p}.$ The
expression for composite velocity $\mathbf{V}$ is given by%
\begin{equation}
\mathbf{V}=\frac{\mathbf{V}_{e}+N_{p}\mathbf{\mathbf{V}_{p}}}{1+N_{p}}.
\label{vv}
\end{equation}%
The composite velocity can also be written as%
\begin{equation}
\mathbf{V}=f_{3}\mathbf{\mathbf{\nabla }\times \mathbf{\nabla }\times 
\mathbf{\mathbf{B}-}}f_{2}\mathbf{\mathbf{\mathbf{\nabla }\times \mathbf{B+}}%
}f_{1}\mathbf{\mathbf{\mathbf{B,}}}  \label{cv}
\end{equation}%
where $f_{3}=(e_{3}+p_{3}N_{p})(1+N_{p})^{-1}$, $%
f_{2}=(e_{2}+p_{2}N_{p})(1+N_{p})^{-1}$ and $%
f_{1}=(e_{1}+p_{1}N_{p})(1+N_{p})^{-1}$. It is clear from Eq. (\ref{cv})
that there exist a strong coupling field and flow which lead to
self-organization of thermally relativistic plasma.

\section{Ideal invariants for TB state}

Knowledge of ideal invariants play an important role to describe the process
of self-organization. Hence it is necessary to look for the constants of
motion. The equation (\ref{ve}) governing the evolution of vorticities
yields the conserved physical quantities known as generalized helicities for
plasma species. The generalized helicities of electron and positron species
are given by%
\begin{equation}
h_{e}=\frac{1}{2}\int \left( \mathbf{\Omega }_{e}\cdot(curl)^{-1}%
\mathbf{\Omega }_{e}\right) dx^{3},
\end{equation}%
\begin{equation}
h_{p}=\frac{1}{2}\int \left( \mathbf{\Omega }_{p}\cdot(curl)^{-1}%
\mathbf{\Omega }_{p}\right) dx^{3},
\end{equation}%
where $h_{e}$ and $h_{p}$ are generalized helicities of electron and
positron fluids respectively. Apart from generalized helicities, the
magnetofluid energy ($\mathcal{E}$) is conserved as well, and can be
expressed as%
\begin{equation}
\mathcal{E}=\frac{1}{2}\int \left[ G\left( V_{e}^{2}+N_{p}V_{p}^{2}\right)
+B^{2}\right] dx^{3}.
\end{equation}%
Hence there exist three ideal invariants namely generalized helicities of
electrons \& positrons and magnetofluid energy. Therefore, in a plasma
consisting of $N$ dynamic species, there will be $N+1$ ideal invariants \cite%
{Mahajan2015}. In order to build the constrained variational principle, it
is necessary to make the implication that generalized helicities are most
robust to dissipation than magnetofluid energy. The functional to be
minimized for plasma system can be written as%
\begin{equation}
\delta \left( \mathcal{E-}\lambda _{1}h_{e}-\lambda _{2}h_{p}\right) =0,
\end{equation}%
where $\lambda _{1}=1/aG$ and $\lambda _{2}=1/bG$ serve as Lagrange
multipliers in the equation. Considering the independent variations of $%
\mathbf{V}_{e}$, $\mathbf{V}_{p}$ and $\mathbf{A}$, and equating the
coefficients of $\delta \mathbf{V}_{e}$, $\delta \mathbf{V}_{p}$ and $\delta 
\mathbf{A}$ on both sides of the above equation and after some algebraic
manipulation, the Eq. (\ref{tcb}) representing the equilibrium state can be
retrieved.

\section{Characteristics of TB state}

As the curl operators are commutative, hence Eq. (\ref{tcb}) can be written
as superposition of three linear Beltrami fields $\mathbf{F}_{\alpha }$. The
Beltrami fields $\mathbf{F}_{\alpha }$ satisfy the relation $\mathbf{\nabla }%
\times \mathbf{F}_{\alpha }=\mu _{\alpha }\mathbf{F}_{\alpha }$, where $\mu
_{\alpha }$ are the eigenvalues of the curl operator \cite{Yoshida1990}. The
examples of the Beltrami fields ($\mathbf{F}_{\alpha }$) are
Chandrasekhar-Kendall functions in cylindrical geometry \cite%
{Chandrasekhar1956} and Arnold-Beltrami-Childress (ABC) flow in slab
geometry \cite{Yoshida1999}. Introducing the eigenvalues (scale parameters),
Eq. (\ref{tcb}) can be written as%
\begin{equation}
(\mathbf{\nabla }\times -\mu _{1})(\mathbf{\nabla }\times -\mu _{2})(\mathbf{%
\nabla }\times -\mu _{3})\mathbf{B}=0,  \label{ctcb}
\end{equation}%
where $\mu _{1}$, $\mu _{2}$ and $\mu _{3}$ are the eigenvalues of the curl
operator and dimensionally they are inverse of length. The relations between
the scale parameters and coefficients of the TB Eq. (\ref{tcb}) are as
follows%
\begin{eqnarray}
k_{1} &=&\mu _{_{1}}\mu _{_{2}}\mu _{_{3}}, \\
k_{2} &=&\mu _{_{1}}\mu _{_{2}}+\mu _{_{1}}\mu _{_{3}}+\mu _{_{2}}\mu
_{_{3}}, \\
k_{3} &=&\mu _{_{1}}+\mu _{_{2}}+\mu _{_{3}}.
\end{eqnarray}%
One can convert Eq. (\ref{ctcb}) into a cubic equation as given below%
\begin{equation}
\mu ^{3}-k_{3}\mu ^{2}+k_{2}\mu -k_{1}=0.  \label{eve}
\end{equation}%
The roots of Eq. (\ref{eve}) are given below as%
\begin{equation}
\mu _{1}=\frac{a+b}{3}+S+T,  \label{root1}
\end{equation}%
\begin{equation}
\mu _{2}=\frac{a+b}{3}-\frac{1}{2}\left( S+T\right) +\frac{i\sqrt{3}}{2}%
\left( S-T\right) ,  \label{root2}
\end{equation}%
\begin{equation}
\mu _{3}=\frac{a+b}{3}-\frac{1}{2}\left( S+T\right) -\frac{i\sqrt{3}}{2}%
\left( S-T\right) ,  \label{root3}
\end{equation}%
where $S=[\left( P/2\right) -\sqrt{\left( P^{2}/4\right) +\left(
Q^{3}/27\right) }]^{1/3},$ $T=[\left( P/2\right) +\sqrt{\left(
P^{2}/4\right) +\left( Q^{3}/27\right) }]^{1/3},\ P=[\left( a-2b\right)
\left( abG+2a^{2}G-b^{2}G-9\right) +9N_{p}\left( 2a-b\right) ]/27G$ and $%
Q=(3-a^{2}G+abG-b^{2}G+3N_{p})/3G$.\ A comprehensive analysis of Eq. (\ref%
{eve}) can be done with the help of the discriminant $\mathit{D}$ as given
below%
\begin{equation}
\mathit{D}=\left(
4d_{2}^{3}+4G^{2}d_{1}^{3}d_{3}+18Gd_{1}d_{2}d_{3}+27Gd_{3}-Gd_{1}^{2}d_{2}^{2}\right) G^{-3},
\label{disc}
\end{equation}%
where $d_{1}=a+b$, $d_{2}=1+Gab+N_{p}$ and $d_{3}=b+aN_{p}$. The nature of
roots can be determined from the value of $\mathit{D}$. When $\mathit{D}=0$,
all the scale parameters are real and at least two are equal; when $\mathit{D%
}>0$, one root is real and the other two are complex, and if $\mathit{D}<0$,
all the eigenvalues are real and distinct.

\begin{figure}[h]
\centering
\includegraphics{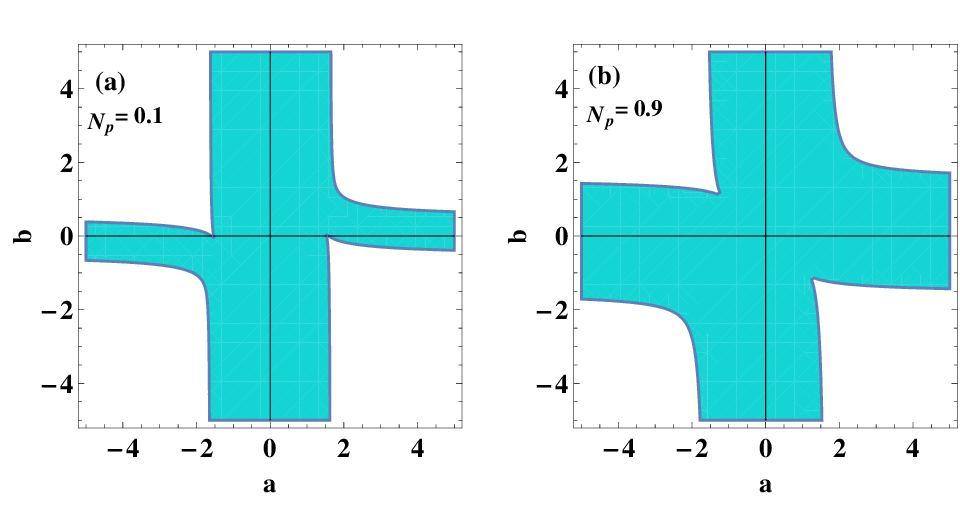}
\caption{Character of the eigenvalues of cubic equation as function of Beltrami
parameters $a$ and $b$ for \ $G=1.5$ and $N_{p}=0.1$ and $N_{p}=0.9$. In the
colored region, the eigenvalues are complex.}
\label{Fig1}
\end{figure}

\begin{figure}[h]
\centering
\includegraphics{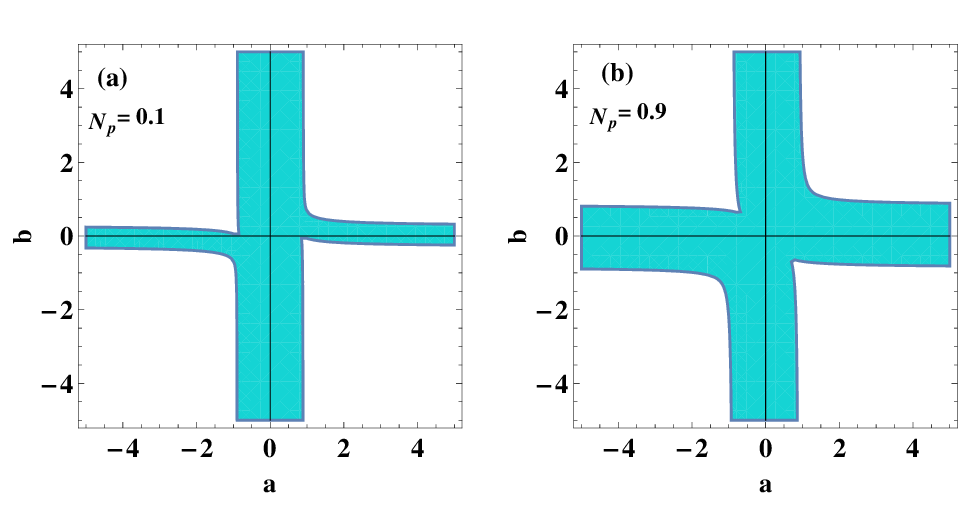}
\caption{Character of the
eigenvalues of cubic equation as function of Beltrami parameters $a$ and $b$
for \ $G=5.0$ and $N_{p}$ \ is $0.1$ and $0.9$. In the colored region, the
eigenvalues are complex.}
\label{Fig2}
\end{figure}

Figures (\ref{Fig1}-\ref{Fig2}) show the character of scale parameters as a
function of Beltrami parameters $a$ and $b$ for the lower and higher
positron densities $N_{p}$ for a fixed value of thermal energy $G\ $of
plasma species. The colored regions of the plot show two complex and one
real eigenvalues while all the eigenvalues are real in the transparent
region. The value of relativistic temperature $G$ is taken to be 1.5 in Fig.
(\ref{Fig1}) while it is set at 5 in Fig. (\ref{Fig2}). In both the cases,
it is evident that at higher positron densities, the complex roots are
increased. That is, on increasing the positron density, some of the real
roots are transformed to complex ones. Consequently, the positron flow
decreases and self-organized vortices become diamagnetic. On the other hand
for a given value of positron density, an increase in the thermal energy
changes some of the complex roots into real roots. The self-organized
vortices become paramagnetic out of diamagnetic. In this case, electrons
flow is increased and generalized vorticity of electrons becomes lesser than
that of electrons flow. 

\begin{figure}[h]
\centering
\includegraphics{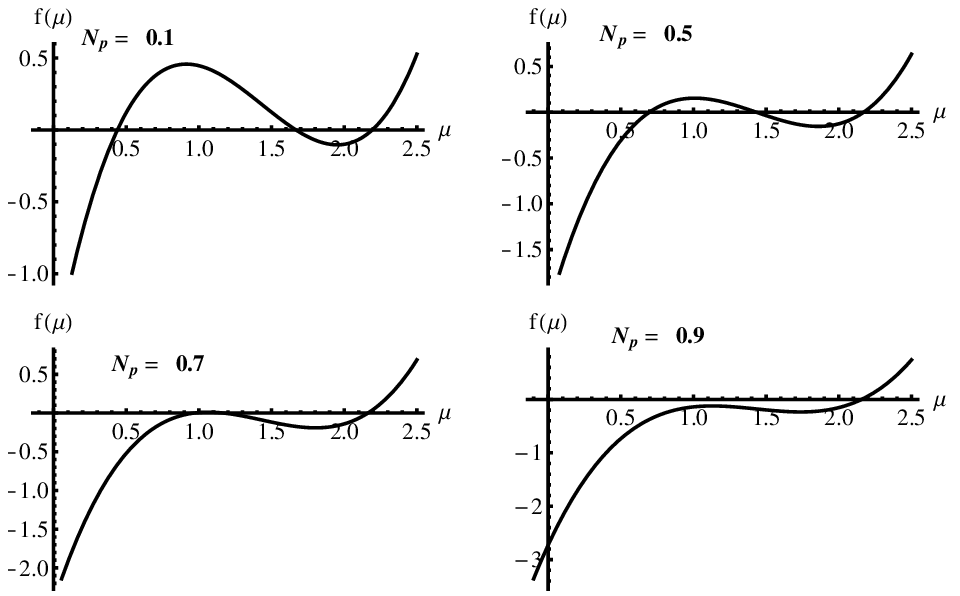}
\caption{Character of
eigenvalues for different values of positron density for $a=2.1$, $b=2.2$ and $G=1.5$.}
\label{Fig3}
\end{figure}

Now we investigate the effect of relativistic temperature and
positron density on the self-organized structures. Fig. (\ref{Fig3}) shows
the character (real or complex roots of cubic function $f\left( \mu \right)
=\mu ^{3}-k_{3}\mu ^{2}+k_{2}\mu -k_{1}$) for different values of positron
density $N_{p}$\ when\textbf{\ }$a=2.1$\textbf{, }$b=2.2$\textbf{\ }and%
\textbf{\ }$G=1.5$\textbf{.} Generally, the real roots give the paramagnetic
From the plot, it is clear that in the slightly relativistic regime when
positron density is very small ($N_{p}=0.1$), all the eigenvalues are real
and the one of the eigenvalues is approximately the order of Beltrami
parameter $b$. The values of scale parameters are $\mu _{1}=2.1928$, $\mu
_{2}=1.6678$ and $\mu _{3}=0.4393$. But for positron density equal to $0.9$
or higher than this, one large scale parameter remains real while the other
two small scale parameters transform to the complex conjugate pair and their
values are $\mu _{1}=2.1568$, $\mu _{2}=1.0716+0.3404i$ and $\mu
_{3}=1.0716-0.3404i$. This shows that for given values of Beltrami
parameters and thermal energy, the scale parameters become complex at higher
positron density. 

\begin{figure}[h]
\centering
\includegraphics{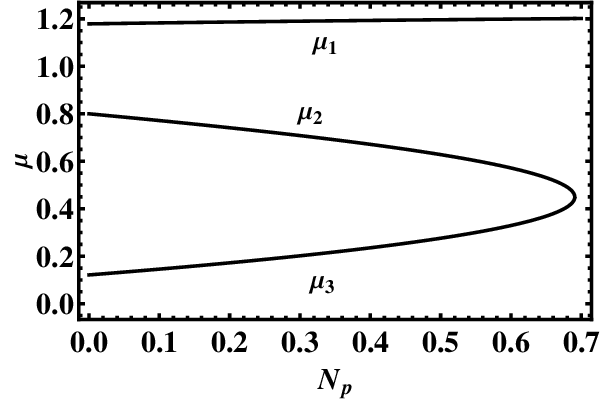}
\caption{Variation in scale-parameters as a function of positron density $N_{p}$
for $a=1.3$, $b=0.8$ and $G=7.0$.}
\label{Fig4}
\end{figure}

The impact
of positron density on the size of self-organized vortices for the given
values of Beltrami parameters and highly relativistic temperature is
illustrated in Fig. (\ref{Fig4}). For $a=1.3$, $b=0.8$ and $G=7.0$, the plot
shows that for lower positron density, all the roots are real, distinct and
disparate which depicts the formation of multiscale structures\textbf{.} It
is also evident from the Fig. (\ref{Fig4}) that on increasing the positron
density, the value of scale parameter $\mu _{1}$ slightly increases, $\mu
_{2}$ decreases while $\mu _{3}$ increases. When the positron density $%
N_{p}>0.7,$ the two small scale parameters $\mu _{2}\ $and $\mu _{3}$ become
complex while the large scale parameter $\mu _{1}$ remains real. Figs. (\ref%
{Fig3}-\ref{Fig4}) demonstrate that the positron density has a significant
effect on the TB state for a variety of Beltrami parameters and relativistic
temperatures. By changing the positron density, one may vary the nature and
size of self-organized structures. 

\begin{figure}[h]
\centering
\includegraphics{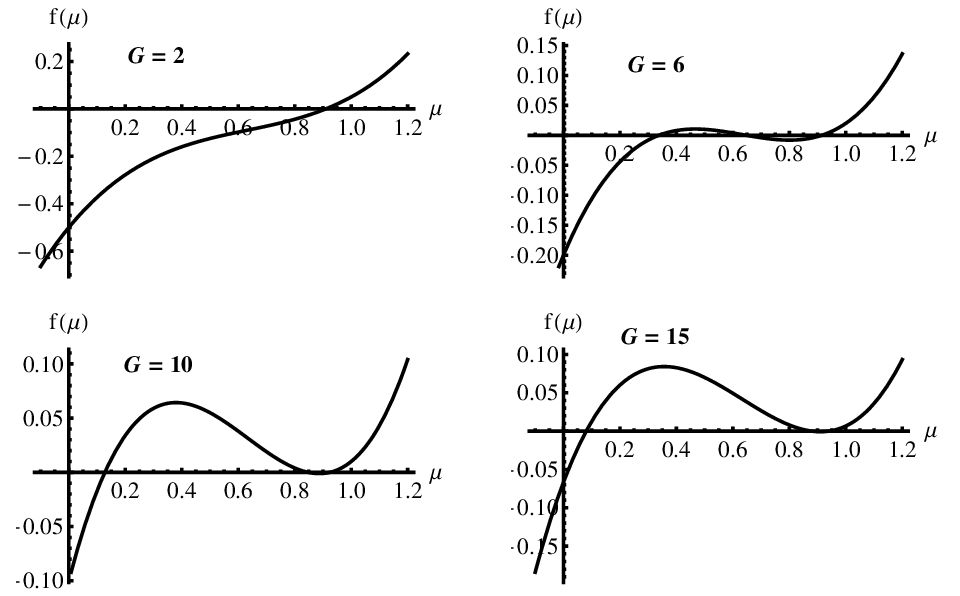}
\caption{Character of eigenvalues for different values of thermal energy $G$
for $a=1.0$, $b=0.9$ and $N_{p}=0.1$.}
\label{Fig5}
\end{figure}

\begin{figure}[h]
\centering
\includegraphics{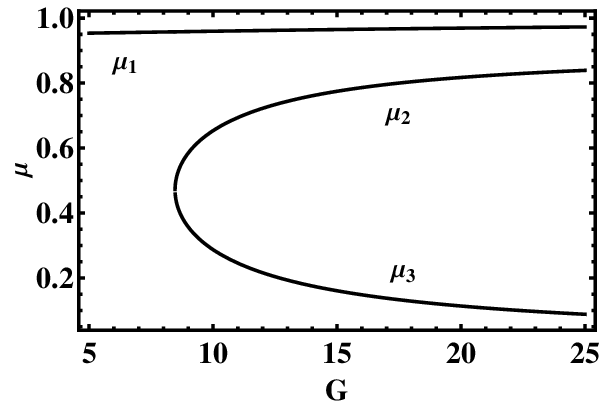}
\caption{Variation in the eigenvalues as a function of thermal
energy $G$ for $a=1.0$, $b=0.9$ and $N_{p}=0.9$.}
\label{Fig6}
\end{figure}

Figure (\ref{Fig5}) shows the character of eigenvalues for
different values of thermal energy $G$\ in case of $a=1.0$\textbf{, }$b=0.9$%
\textbf{\ }and\textbf{\ }$N_{p}=0.1$\textbf{.} The graph depicts that for
lower relativistic temperature $G=2$, one eigenvalue is real ($\mu
_{1}=0.9106$) and other two are complex ($\mu _{2}=0.4946+0.5516i$ and $\mu
_{3}=0.4946+0.5516i$). But for ultra-relativistic temperature $G=15,\ $all
the eigenvalues are real, distinct and disparate and given by $\mu
_{1}=0.9397$, $\mu _{2}=0.8795$ and $\mu _{3}=0.0806$. This depicts that
when the relativistic temperature increases, eigenvalues change their
character and complex roots are transformed to real ones. It also shows that
one scale parameter approaches to zero with an increase in relativistic
temperature which provides the possibility of formation of macroscopic
structure of the order of system size.  
Figure (\ref{Fig6}) illustrates the variation in size of
self-organized vortices as a function of relativistic temperature $G$\ for
given values of Beltrami parameters $a=1.0$\ and $b=0.9$\ and positron
density\textbf{\ }$N_{p}=0.9$. The graph depicts that for lower relativistic
temperature, one scale parameter is real and other two are complex
conjugate. As the relativistic temperature increases, all the roots become
real, distinct and separate when $G\geq 8.5$. It is interesting to note that
the scale parameter $\mu _{2}$ decrease with an increase in thermal energy
and approaches to zero at ultra-relativistic temperature. On the other hand,
values of scale parameters $\mu _{1}$ and $\mu _{3}$ keep on increasing with
an increase in relativistic temperature. The graph also shows that at
ultra-relativistic temperature, one of self-organized structures is
macroscopic corresponding to $\mu _{2}\simeq 0,$ whereas the other two
structures are smaller ones of the order of skin depth. At
ultra-relativistic temperature, the thermal energy satisfies the condition $%
G>>\left\vert ab\right\vert ^{-1}$. The eigenvalues corresponding to this
condition are given by $\mu _{1}\approx a,$ $\mu _{2}\approx 0$ and $\mu
_{3}\approx b.$ The large scale structure equal to system size corresponds
to $\mu _{2}\approx 0$. This shows the possibility of creating large scale
structures in an ultra-relativistic EPI plasma. 
The Beltrami parameters, relativistic temperature and positron
density play a vital role in the formation of self-organized multiscale
structures. Let us consider some special conditions and their impact on the
nature and values of scale parameters. When the ratios of generalized
vorticities of electron and positron species to the respective flows are
taken equal ($a=b$), then one eigenvalue is real and is equal to the
Beltrami parameter $\mu _{1}=a$ and other two are given by $\mu
_{2,3}=\left( a\sqrt{G}\pm \sqrt{a^{2}G-4\left( N_{p}+1\right) }\right) /2%
\sqrt{G}$. For $a^{2}=4\left( N_{p}+1\right) /G$ , all the eigenvalues are
real and two of them are equal. The eigenvalues are as follows: $\mu _{1}=%
\sqrt{4\left( N_{p}+1\right) /G}$ and $\mu _{2,3}=\sqrt{\left(
N_{p}+1\right) /G}$. In case of $a^{2}>4\left( N_{p}+1\right) /G$, all the
roots are real and distinct while for $a^{2}<4\left( N_{p}+1\right) /G$, one
root is real and other two are complex conjugate pair. For instance when $%
G=5.0$ and $N_{p}=0.9,$ roots are real for $a\geq 1.24.$ When $a=1.24,$ the
roots are $\mu _{1}=1.24$, $\mu _{2}=0.5536$ and $\mu _{3}=0.6863$. From
Fig. 1 and 2, it is clear that when the generalized vorticity of either
electron or positron species vanishes ($a$ or $b=0$), one of the roots is
real and other two are complex conjugate. When the generalized vorticities
of both species vanish ($a=b=0$), then the flow vorticities are aligned to
magnetic field. In this scenario, one of the scale parameters become zero ($%
\mu _{1}=0$) while the other two are complex conjugate ($\mu _{2,3}=\pm i%
\sqrt{\left( 1+N_{p}\right) /G}$) and the TB state is transformed to DB
state. The imaginary eigenvalues show perfect diamagnetic behavior and it is
completely dependent on positron density and relativistic temperature \cite%
{Mahajan2008}. When positron density is negligible $\left( N_{p}\simeq
0\right) ,$ the scale parameters are $\mu _{1}=0$ and $\mu _{2,3}=\pm i\sqrt{%
1/G}$. In case of pure electron-positron plasma ($N_{p}=1$), the eigenvalues
become $\mu _{1}=0$ and $\mu _{2,3}=\pm i\sqrt{2/G}$. For negligible
positron density $N_{p}\simeq 0$, the eigenvalues are given by $\mu _{1}=b$
and $\mu _{2,3}=a/2\pm \sqrt{\left( a^{2}G-4\right) /4G}$. The roots are
real when $a>2/\sqrt{G}$ and $b>a/2+\sqrt{\left( a^{2}G-4\right) /4G}$. The
Beltrami parameters and thermal energy can be related to eigenvalues and
given by $b=\mu _{1}$, $a=\mu _{2}+\mu _{3}$ and $G=\left( \mu _{2}\mu
_{3}\right) ^{-1}.$ For $G=7.0$, $N_{p}\simeq 0$, $a=0.85$ and $b=0.73,$ the
roots are $\mu _{1}=0.73$, $\mu _{2}=0.2272$ and $\mu _{3}=0.6287$.

\section{Impact of positron density and thermal energy on self-organized
structures}

\begin{figure}[h]
\centering
\includegraphics{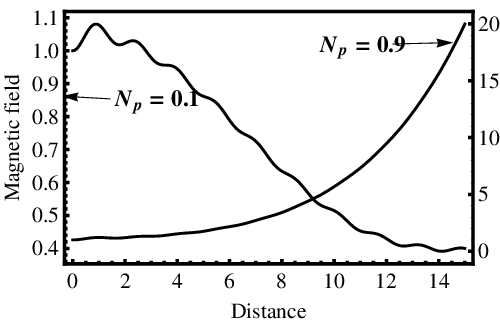}
\caption{Variation of the magnetic field for $N_{p}=0.1$ (left vertical axis) and $
N_{p}=0.9$ (right vertical axis) for $a=4.3$, $b_{p}=0.3$ and $G=10$.}
\label{Fig7}
\end{figure}

The thermally relativistic EPI plasmas are ubiquitous in nature and can be
produced in intense ultra-short laser beam experiments. Therefore the
current analysis has significance in understanding the laboratory and
astrophysical plasmas.\textbf{\ }For instance, the pulsar magnetospheres are
composed mostly of thermally relativistic secondary electron-positron
plasma, with small quantities of ions present in certain cases. This plasma
has the potential to influence the radiation generated in the inner area of
the magnetosphere as well as at the stellar surface, among other things. In
order to interpret observations, it is essential to understand the
characteristics of the pulsar magnetosphere plasma. For pulsar
magnetospheric plasma, the electron density is $n_{e}=10^{6}$cm$^{-3}\ $%
(corresponding skin depth is $5.31\times 10^{2}$cm)\ at a distance of $10^{8}
$cm\ from pulsar surface \cite{Melrose1978,Michel1982,Soto2010,Lazarus2012}.
By ignoring the differential rotation of pulsar magnetosphere one can
consider cylindrical geometry.\textbf{\ }The analytical solution of magnetic
field and flow for a one dimensional cylindrical geometry can be written as,%
\begin{equation}
B=\sum\limits_{\alpha =1}^{3}C_{\alpha }\left( 
\begin{array}{c}
0 \\ 
J_{1}(\mu _{\alpha }r) \\ 
J_{0}(\mu _{\alpha }r)%
\end{array}%
\right) ,  \label{sol}
\end{equation}%
where $C_{\alpha }$ are constants and can be determined by boundary
conditions and 
\begin{equation}
V=\sum\limits_{\alpha =1}^{3}C_{\alpha }j_{\alpha }\left( 
\begin{array}{c}
0 \\ 
J_{1}(\mu _{\alpha }r) \\ 
J_{0}(\mu _{\alpha }r)%
\end{array}%
\right) ,  \label{bfv}
\end{equation}%
where $j_{1}=\mu _{1}^{2}f_{3}-\mu _{1}f_{2}+f_{1}$, $j_{2}=\mu
_{2}^{2}f_{3}-\mu _{2}f_{2}+f_{1}$ and $j_{3}=\mu _{3}^{2}f_{3}-\mu
_{3}f_{2}+f_{1}$. To calculate the values of constants $C_{\alpha }$, we use
the following boundary conditions, $B_{z}|_{r=0}=g$, $B_{\theta
}|_{_{r=r_{0}}}=h$, and $\left\vert \nabla \times B_{_{\theta }}\right\vert
_{r=r_{0}}=s$, where $g=\sum\limits_{\alpha =1}^{3}C_{\alpha }$, $%
h=\sum\limits_{\alpha =1}^{3}C_{\alpha }J_{1}(\mu _{\alpha }r_{0})$, $%
s=\sum\limits_{\alpha =1}^{3}\mu _{\alpha }C_{\alpha }$ and $r_{0}$ are
arbitrary and real valued. The expressions for constants $C_{1}$, $C_{2}$
and $C_{3}$ are,%
\begin{equation*}
C_{1}=\frac{\left( g\mu _{3}-s\right) J_{1}(\mu _{2}r_{0})+\left( s-g\mu
_{2}\right) J_{1}(\mu _{3}r_{0})+\left( \mu _{2}-\mu _{3}\right) h}{%
J_{1}\left( r_{0}\mu _{3}\right) \left( \mu _{1}-\mu _{2}\right)
+J_{1}\left( r_{0}\mu _{1}\right) \left( \mu _{2}-\mu _{3}\right)
+J_{1}\left( r_{0}\mu _{2}\right) \left( \mu _{3}-\mu _{1}\right) },
\end{equation*}%
\begin{equation*}
C_{2}=\frac{\left( g\mu _{1}-s\right) J_{1}(\mu _{3}r_{0})+\left( s-g\mu
_{3}\right) J_{1}(\mu _{1}r_{0})+\left( \mu _{3}-\mu _{1}\right) h}{%
J_{1}\left( r_{0}\mu _{3}\right) \left( \mu _{1}-\mu _{2}\right)
+J_{1}\left( r_{0}\mu _{1}\right) \left( \mu _{2}-\mu _{3}\right)
+J_{1}\left( r_{0}\mu _{2}\right) \left( \mu _{3}-\mu _{1}\right) },
\end{equation*}%
\begin{equation*}
C_{3}=\frac{\left( g\mu _{2}-s\right) J_{1}(\mu _{1}r_{0})+\left( s-g\mu
_{1}\right) J_{1}(\mu _{2}r_{0})+\left( \mu _{1}-\mu _{2}\right) h}{%
J_{1}\left( r_{0}\mu _{3}\right) \left( \mu _{1}-\mu _{2}\right)
+J_{1}\left( r_{0}\mu _{1}\right) \left( \mu _{2}-\mu _{3}\right)
+J_{1}\left( r_{0}\mu _{2}\right) \left( \mu _{3}-\mu _{1}\right) }.
\end{equation*}%
The magnetic field profile is shown in Fig. (\ref{Fig7}) for $N_{p}=0.1$ and 
$N_{p}=0.9$. The relativistic temperature is taken as $G=10$ and the values
of other parameters are as follows: $a=4.3$, $b=0.3$, $g=1.0$ and $h=s=0$.
When the positron density $N_{p}$ is $0.9$; one scale parameter is real,
while the other two are complex conjugate pair and their values are $\mu
_{1}=4.2767,$ $\mu _{2}=0.1616+0.2671i$ and $\mu _{3}=0.1616-0.2671i$. When
the scale parameters are complex, the magnetic field is maximum on the
system's boundary, so the magnetic structure shows diamagnetic behavior \cite%
{Mahajan1998,Mahajan2008}. In Fig. (\ref{Fig7}) when the positron density $%
N_{p}$ is $0.1$; the magnetic field is minimum on the system's boundary, so
the magnetic structure shows paramagnetic behavior. The scale parameters are
real and distinct ($\mu _{1}=4.2766$, $\mu _{2}=0.0664$ and $\mu _{3}=0.2569$%
). It is important to note that one of the scale parameters ($\mu _{1}=4.2766
$) remains constant for the both the densities $N_{p}=0.1$ and $N_{p}=0.9$.
While on the other hand the complex roots at $N_{p}=0.9$ are transformed to
real roots at $N_{p}=0.1$.

\begin{figure}[h]
\centering
\includegraphics{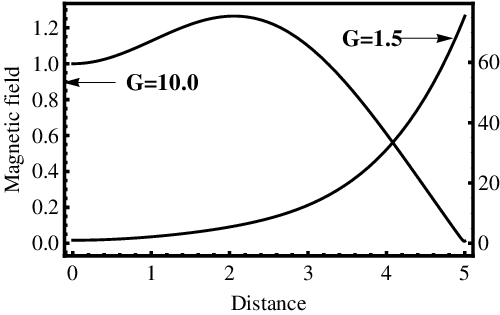}
\caption{Variation of the magnetic
field for $G=10$ (left vertical axis) and $G=1.5$ (right vertical axis) for $%
a=1.1$, $b=1.4$ and $N_{p}=0.9$.}
\label{Fig8}
\end{figure}

Fig. (\ref{Fig8}) shows the variation
in the magnetic field for lower and higher relativistic temperature $G$ of
plasma species when positron density $N_{p}$ is $0.9$\ and the Beltrami
parameters are $a=1.1$ and $b=1.4$ while the boundary conditions are $g=1$
and $h=s=0$. When the thermal energy of plasma species is $G=1.5$; one root
is real, while the other two are complex conjugate pair, and their values
are $\mu _{1}=1.2797,$ $\mu _{2}=0.6101+0.9342i$ and $\mu _{3}=0.6101-0.9342i
$. Corresponding to these eigenvalues the magnetic field is maximum on the
system's boundary as compared to interior of the plasma, so the
self-organized structure shows diamagnetic behavior \cite%
{Mahajan1998,Mahajan2008}. When the relativistic temperature $G$ is
increased to $10.0$; the scale parameters are real and distinct ($\mu
_{1}=0.1830$, $\mu _{2}=0.9684$ and $\mu _{3}=1.3486$). For these real
eigenvalues, the magnetic field is minimum on the system's boundary, so the
magnetic structure shows paramagnetic behavior. It is clear from Fig. (\ref%
{Fig8}) that for lower relativistic temperature, magnetic structures are
diamagnetic but for higher thermal energies it shifts to paramagnetic
structure and eigenvalues change their nature. The decaying magnetic field
or the presence of paramagnetic structures indicate the presence of magnetic
reconnect. Among the major energy conversion mechanisms in plasmas, magnetic
reconnection is one of the most significant. It converts magnetic field
energy into plasma kinetic energy and high-energy particles \cite%
{Sakai2001,Mahajan2018a,Lazarian2020}. It is possible to think of these
paramagnetic structures as a source of plasma heating and streaming
particles in the magnetosphere, and they may be regarded to be such.

\section{Dynamo Mechanisms}

The self-organized TB state is characterized by three scale parameters\ $\mu
_{j}$ which are the eigenvalues of the curl operators and characterize the
reciprocal of length scales on which the magnetic field and velocity change
significantly. The scale parameters can have quite different values when the
Beltrami parameters $a$ and $b$ are varied for a given temperature and
densities of the species. When length scales are separated significantly,
the magnetic field and the associated velocity may have an appreciable
difference. The magnetic field may get dominated by the long scale field $%
\left\vert \mu _{j}\right\vert \ll 1$ while the flow may have a large
component varying on the short scale $\left\vert \mu _{j}\right\vert \gg 1$.
The opposite phenomenon may also happens.

\begin{figure}[h]
\centering
\includegraphics{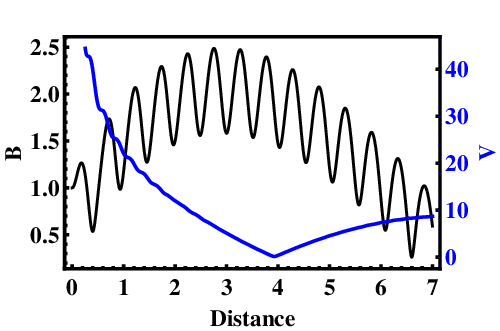}
\caption{Jittery magnetic field (right vertical axis) coupled
with smooth flow (left vertical axis) for $a=12.7$, $b=11.9$, $N_{p}=0.9,$ $
G=7.$}
\label{fig9new}
\end{figure}

Let us examine two different regimes of parameters. First, we consider the
case when both $\left\vert a\right\vert $ and $\left\vert b\right\vert $ are
relatively large and $a\approx b$. In this case, the flow dominates the
dynamics while the magnetic field is relatively small as depicted in Fig. (%
\ref{fig9new}) which is plotted for $a=12.7$, $b=11.9$, $G=7.0$ and $%
N_{p}=0.9$. The corresponding scale parameters are $\mu _{1}=0.022,$ $\mu
_{2}=11.88$ and $\mu _{3}=12.68\ $while the boundary conditions are taken as 
$g=1.0,\ h=0.1$ and $s=0.5$. For these eigenvalues and plasma parameters,
the relation between flow and field using equations (\ref{sol}-\ref{bfv}) is 
$V=14.66B$, which confirms that $V\gg B$. As a very small $\left\vert
B\right\vert $ is associated with a strong flow, one can say that magnetic
field is being generated by induction\ effect. These conditions are relevant
to the fast dynamo \cite{Tanner2003,Iqbal2012PoP}.

\begin{figure}[h]
\centering
\includegraphics{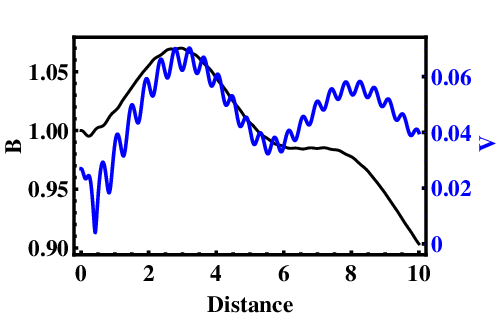}
\caption{Smooth magnetic field (right vertical axis) coupled with
jittery flow (left vertical axis) for $a=15.0$, $b=1.4$, $N_{p}=0.9$, $G=7$.}
\label{fig10new}
\end{figure}

Next, we suppose that $a=15.0$ and $b=1.4$ while all other parameters are
taken same as that of Fig. (\ref{fig9new}).The corresponding scale
parameters in this case are $\mu _{1}=0.1092$, $\mu _{2}=1.3003$ and $\mu
_{3}=14.99$. For these plasma parameters, Fig. (\ref{fig10new}) shows an
approximately smooth profile of magnetic field while the flow is jittery. In
contrast to the case displayed in Fig. (\ref{fig9new}), the plot depicts
that magnetic field is stronger as compared to flow and it varies on a long
scale while on the other hand, flow is varying on a short scale. This
scenario provides the generic turbulent dynamo when an ordered magnetic
field is created out of a complex flow \cite{Vainshtein1991,Iqbal2012PoP}.
Also for this case the relation between flow and field comes out to be $%
V=0.05B$, which confirms that $V\ll B$.

The dynamo processes described here are highly feasible contenders as well
as a strong indication for producing large scale magnetic fields and fast
outflows in a myriad of astrophysical environments. 

The above discussion shows the
possibility of creating three self-organized vortices when a thermally
relativistic EPI plasma attains its steady state under appropriate
constraints. No doubt, it is possible to create more complex structures in
this system, only the field and flow vortices which give rise to dynamo
mechanisms are presented in this work.

\section{Summary}

The relaxation of a thermally relativistic electron and positron plama
containing static ions is studied and the impact of relativistic
temperatures and densities of the species is analyzed. The relaxed state is
found to be a TB state which can be regarded as composed of three distinct
Beltrami states. Generally, two of the eigenvalues of TB state are complex
conjugate while the third one is real. It is found that when the positron
density decreases, all the scale parameters become real in both relativistic
and non-relativistic regimes. Likewise, when the relativistic temperature
rises, all the eigenvalues become real. Additionally, it is shown that in
the relaxed state of an ultra-relativistic EPI plasma, a macroscopic
structure of system size can be formed together with two electron scale
structures. Furthermore, it is explored how the positron density and
relativistic temperature control the self-organized structures. At lower
positron density and higher relativistic temperature, the plasma exhibits
paramagnetic state, whereas at higher positron density and lower
relativistic temperature, the plasma exhibits diamagnetic behavior. It shows
that a change in density and relativistic temperature can cause the
conversion of magnetic and kinetic energies in the thermally relativistic
plasmas. A complete analysis of dynamo mechanism requires a detailed
analytical and numerical study, which is beyond the scope of the present
work. However, the relaxed equilibria being the consequence of the coupling
of three Beltrami conditions with three spatial scale lengths allow the
simultaneous existence of two fields which vary on vastly different scales
through their self-consistent coupling. This disparate variation of the
magnetic field and velocity is precisley the required condition for the
dynamo mechanism: the turbulent dynamo in a relatively smooth magnetic field
is generated by a short scale velocity, and the kinematic (fast) dynamo in
which the length scales for the two fields are reversed. It is shown through
graphs that the seeds of both the possibilities are there in the
manifestation of the relaxed equilibria charatcerized by TB state. The
dynamo mechanism appears when one of the scale parameters is very small as
compared to the other ones. The dynamo mechanisms outlined here can produce
large scale magnetic fields and fast outflows in a wide range of
astrophysical settings. The current study will be helpful in understanding
the behavior of thermally relativistic plasmas in the laboratory and
astrophysical environments.

\begin{acknowledgments}
The work of M. Iqbal is funded by Higher Education Commission (HEC),
Pakistan under project No. 20-9408/Punjab/NRPU/R\&D/HEC/2017-18.
\end{acknowledgments}

\end{document}